\documentclass[a4paper,11pt]{article}
\usepackage{indentfirst}
\usepackage{hyperref}
\usepackage{multirow}
\usepackage{latexsym,bm,amsmath,amssymb}
\usepackage{graphicx}
\usepackage{epsfig}
\usepackage{epstopdf}
\usepackage{subfigure}
\usepackage{cite}
\usepackage{color}
\usepackage[top=1in, bottom=1in, left=1.in, right=1.in]{geometry}
\usepackage{appendix}
\usepackage{ulem}
\usepackage{slashed}
\usepackage{enumitem}

\newcommand{\bea}{\begin{eqnarray}}
\newcommand{\eea}{\end{eqnarray}}

\begin{document}
\title{\bf\Large Discussions on the line-shape of the $X(4660)$ resonance}
\author
{\sc
Qin-Fang Cao$^{1}$, Hong-Rong Qi$^{2}$, Yu-Fei Wang$^{1}$, Han-Qing~Zheng$^{1,3}$
\vspace*{0.5cm} \\
$^{1}${\it Department of Physics and State Key Laboratory of Nuclear Physics and Technology,} \\
{\it Peking University,  Beijing 100871, China}\\
$^{2}${\it Department of engineering physics, Tsinghua University, Beijing 100084, China}\\
$^{3}${\it Collaborative Innovation Center of Quantum Matter, Beijing 100871, China}
}
\maketitle
\begin{abstract}
A careful reanalysis is made on $e^+e^-\to X(4660)\to (\Lambda_c\bar{\Lambda}_c)/(\psi'\pi\pi)$ processes, aiming at resolving the apparent conflicts between Belle and BESIII data above $\Lambda_c\bar{\Lambda}_c$ threshold. We use a model containing a Breit-Wigner resonance and $\Lambda_c\bar{\Lambda}_c$ four-point contact interactions, with which the enhancement right above the $\Lambda_c\bar{\Lambda}_c$ threshold is well explained by a virtual pole generated by $\Lambda_c\bar{\Lambda}_c$ attractive final state interaction, located at $M_V=4.566\pm0.007$ GeV. Meanwhile, $X(4660)$ remains to be a typical Breit-Wigner resonance, and is hence of confinement nature. Our analysis strongly suggests the existence of the virtual pole with statistical significance of $4.2$ standard deviation ($\sigma$). Nevertheless, the conclusion crucially depends on the line-shape of cross sections which are of limited statistics, hence we urge new experimental analyses from Belle II, BESIII, and LHCb to settle the issue.
\end{abstract}

Since the discovery of $X(3872)$ in 2003~\cite{Choi:2003ue}, hadronic exotic states, or called ``$XYZ$'' states, open a new window for hadron physics researches. Those states do not match the energy level positions predicted by naive quark models (e.g., the Godfrey-Isgur model in $c\bar{c}$ sector~\cite{Godfrey:1985xj}), and most of them are very narrow in spite of locating above open charm (bottom) thresholds, which have intrigued theorists for recent years. Various models are established to understand such states: for example, modified quark models~\cite{Sultan:2014oua,Barnes:2005pb,Radford:2007vd} which treat them as confining states; ``nonresonance'' interpretations regarding them as branch cut singularities~\cite{Szczepaniak:2015eza}; also, one widely accepted mechanism is so called ``hadronic molecules''~\cite{Guo:2017jvc,Xiao:2016dsx,Xiao:2016wbs,Xiao:2016mon}.

Particularly, in 2007, Belle collaboration observed two structures, dubbed as $X(4360)$ and $X(4660)$, in cross section shape of $e^+e^-\to \gamma_{ISR}\psi'\pi^+\pi^-$~\footnote{In this paper $\psi'$ denotes the $\psi(2S)$ particle. } with quantum number $J^{PC}=1^{--}$~\cite{Wang:2007ea}. This discovery is confirmed later by BABAR collaboration~\cite{Lees:2012pv} and updated observation of Belle~\cite{Wang:2014hta}. Moreover, the investigation of $e^+e^-\to \gamma_{ISR}\Lambda_c\bar{\Lambda}_c$ process by Belle collaboration reveals an ``exotic'' state called $X(4630)$~\cite{Pakhlova:2008vn}. It is believed that $X(4630)$ and $X(4660)$ may in fact be the same state~\cite{Dai:2017fwx,Tanabashi:2018oca,Guo:2010tk}, and various interpretations are proposed, see, e.g., Refs.~\cite{Li:2009zu,Cotugno:2009ys,Guo:2010tk,Guo:2016iej,Liu:2016sip,Gui:2018rvv}.

More recently, a much precise measurement by BESIII collaboration gives the cross sections at four center-of-mass energies for $e^+e^-\to \Lambda_c\bar{\Lambda}_c$ cross section near $\Lambda_c\bar{\Lambda}_c$ threshold~\cite{Ablikim:2017lct}. As shown in Fig.~2 of Ref.~\cite{Ablikim:2017lct} (or the left panel of Fig.~\ref{fig:fitc} in this paper), the BESIII data is questioned to have conflicts with the Belle data~\cite{Dai:2017fwx,Ferroli:2018}: especially the line-shape of BESIII data appears nearly horizontal while that from Belle shows a significant growth. The results from some earlier works are not compatible with BESIII data, see e.g. Fig.~2 of Ref.~\cite{Guo:2010tk}.

\begin{figure*}[htbp]
\centering
\includegraphics[width=0.6\textwidth]{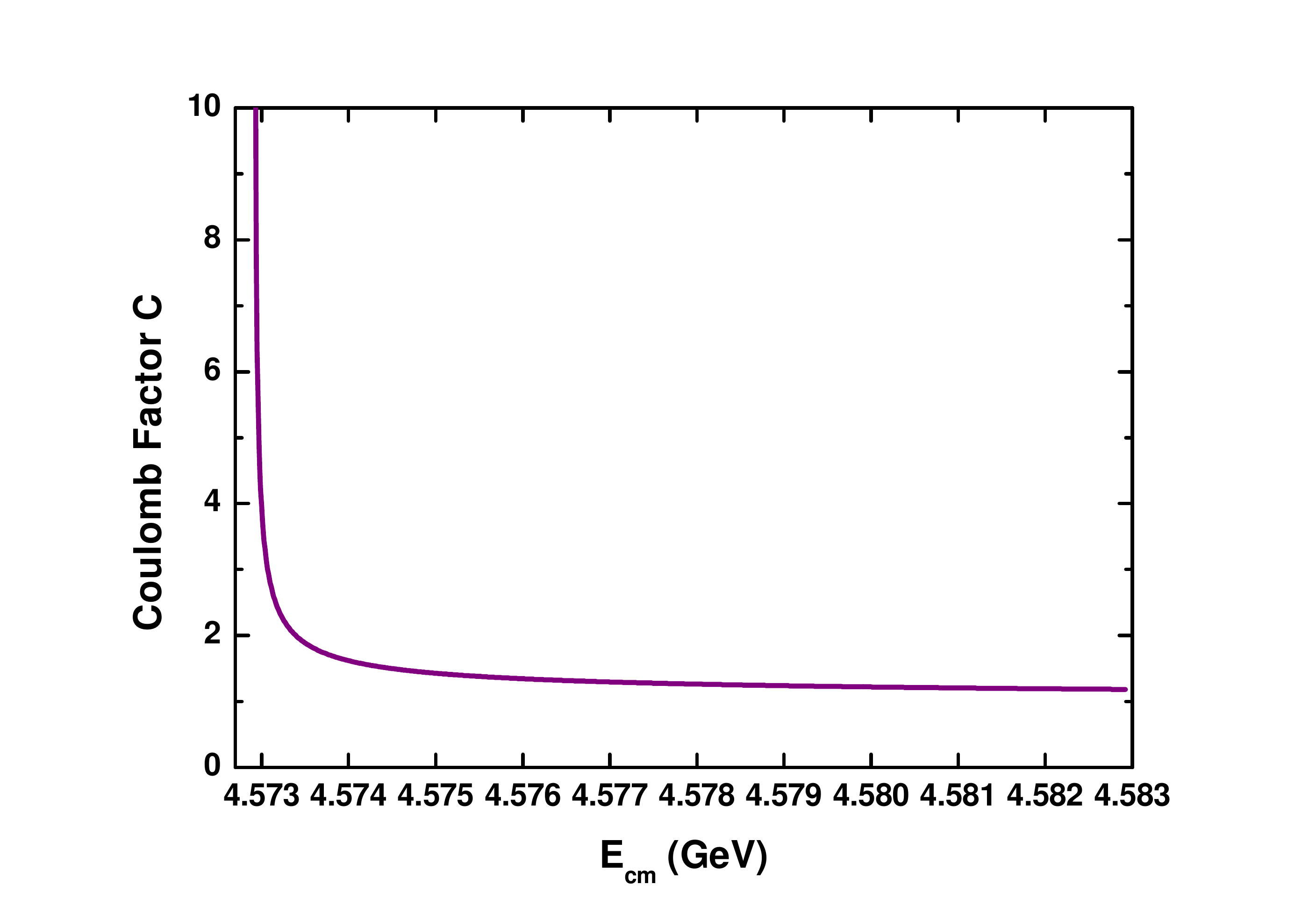}
\caption{\label{fig:CoF} The factor from Coulomb interaction in Eq.~\eqref{CoF}. }
\end{figure*}

One may suspect such odd behavior of the line-shape stems from the threshold enhancement of electromagnetic force, but this can hardly be the actual case. Specifically, the Coulomb interaction is simulated by the well-known Sommerfeld factor:
\begin{equation}\label{CoF}
C(s)=\frac{\pi \alpha m}{k}\left[1-\exp\Big(-\frac{\pi\alpha m}{k}\Big)\right]^{-1}\ ,
\end{equation}
with $\alpha\sim 1/137$ and $k$ being the three-momentum of $\Lambda_c$ and $\bar{\Lambda}_c$ in center-of-mass frame, and $m$ here is the mass of $\Lambda_c$. As plotted in Fig.~\ref{fig:CoF},
the Coulomb interaction indeed provides a spike near threshold, but it soon becomes mild when the center-of-mass energy increases about $1$ MeV. However, the BESIII data indicates the threshold enhancement spreads nearly 30 MeV above threshold. Just as expected, the fit with Sommerfeld factor and $s$ channel $X$ state (to be displayed as Fig.~\ref{fig:Cofit}) cannot match the BESIII data.

This work aims at disentangling this problem: it is suggested that a virtual pole via $\Lambda_c\bar{\Lambda}_c$ contact interactions, in addition to the $X(4660)$ Breit-Wigner resonance, could well explain the odd line-shape which is failed to be described by previous studies.

To begin with, we assume that $X(4630)$ and $X(4660)$ are the same particle and denote it as $X$, with quantum number $^{2s+1}L_J=^3S_1$~\footnote{The $S$-$D$ mixing is omitted since it is suppressed in near-threshold region. For the standard method to calculate amplitudes in $JLS$ basis, see, e.g., Ref.~\cite{Machleidt:1987hj}. }. For $\Lambda_c\bar{\Lambda}_c$ channel, the coupling among $X \Lambda_c\bar{\Lambda}_c$ and the QED transition from a photon to $X$ are introduced as the following:
\begin{equation}\label{intXLam}
\mathcal{L}_{X\Lambda_c}=g_1\bar \Lambda_c\gamma^\mu X_\mu\Lambda_c,\quad \mathcal{L}_{\gamma X}=g_\gamma X^{\mu\nu}F_{\nu\mu};\
\end{equation}
where $X^{\mu\nu}$ and $F_{\nu\mu}$ are the strength tensors of the $X$ and the photon, respectively. The contact interaction between $\Lambda_c\bar{\Lambda}_c$ is
\begin{equation}\label{int4Lam}
\mathcal{L}_{\Lambda_c\bar \Lambda_c}=C_V(\bar\Lambda_c\gamma_\mu\Lambda_c)(\bar\Lambda_c\gamma^\mu\Lambda_c)
+C_A(\bar\Lambda_c\gamma_\mu\gamma^5\Lambda_c)(\bar\Lambda_c\gamma^\mu\gamma^5\Lambda_c),\
\end{equation}
which simulates vector and axial-vector meson exchanges~\footnote{We have considered other types of contact terms, but they cannot fit the data well. }. For $\psi'\pi\pi$ channel, the interactions may be complicated with various Lorentz structures, but only the final two pions with total isospin and angular momentum $IJ=00$ and their final state interaction (FSI) are considered, just like Ref.~\cite{Gong:2016hlt}. Therefore the momentum dependence from derivative couplings of pions can be absorbed into their FSI, leaving the effective vertices as following
\begin{equation}\label{Vfdef}
V_f^{X\psi\pi}\propto g_2\mathcal{A}_{\pi\pi},\quad V_f^{\Lambda_c\psi\pi}\propto g_3\mathcal{A}_{\pi\pi},
\end{equation}
where $g_{2,3}$ are fit parameters and the $\mathcal{A}_{\pi\pi}$ stands for the FSI term between two pions (see e.g. Ref.~\cite{Gong:2016hlt}).

\begin{figure*}[t]
\centering
\includegraphics[width=0.6\textwidth]{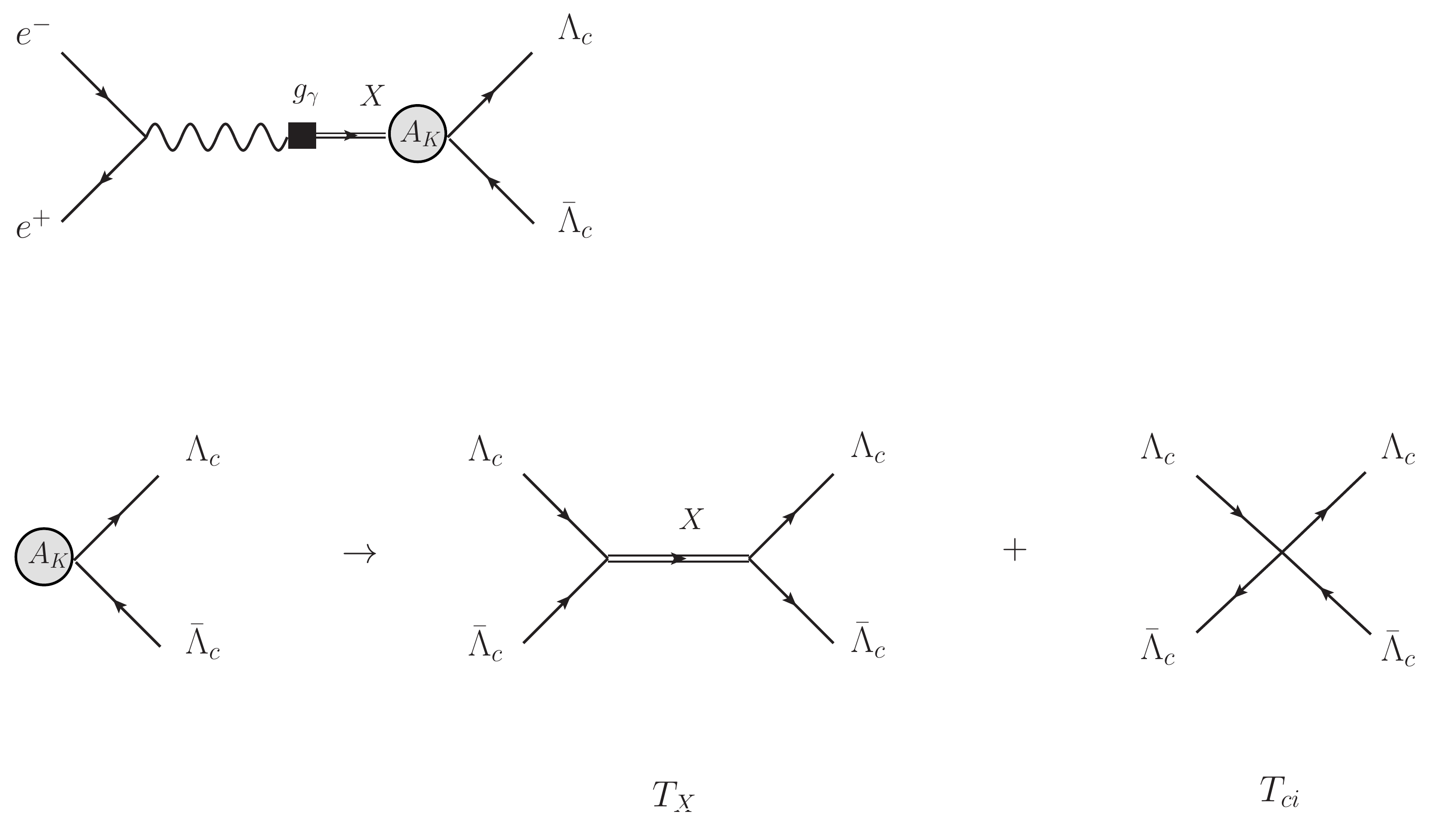}
\caption{\label{fig:KFSlam} The Feynman diagram of $e^+e^-\to X(4660)\to \Lambda_c\bar{\Lambda}_c$. $A_K$ is the $K$ matrix sector in Eq.~\eqref{Km}. }
\end{figure*}

\begin{figure*}[t]
\centering
\includegraphics[width=0.6\textwidth]{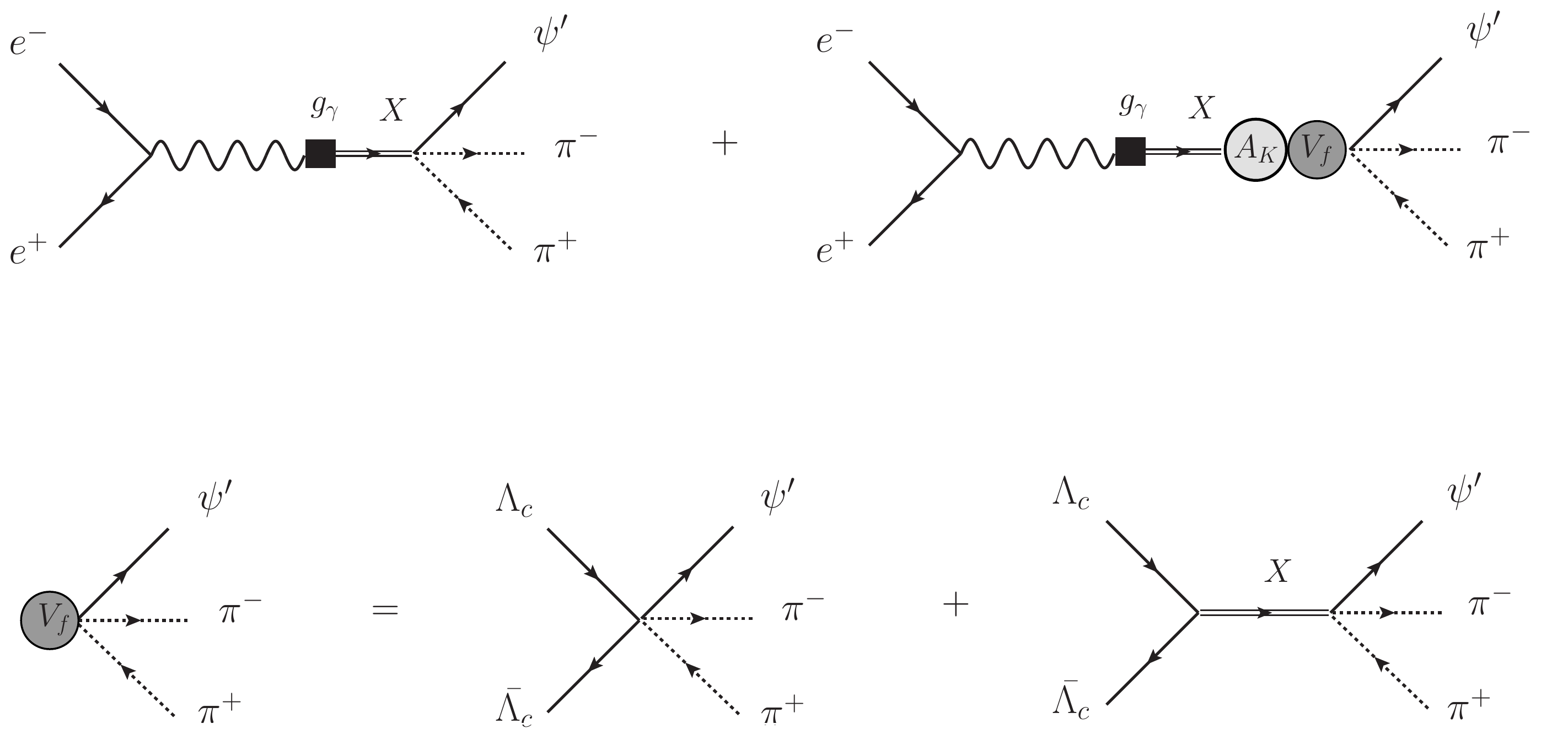}
\caption{\label{fig:KFSpsi} The Feynman diagram of $e^+e^-\to X(4660)\to \psi'\pi\pi$. $A_K$ is the $K$ matrix sector in Eq.~\eqref{Km} (the same as Fig.~\ref{fig:KFSlam}), and $V_f$ is the vertex concerning $\psi'\pi\pi$ final state. }
\end{figure*}

\begin{figure*}[t]
\centering
\includegraphics[width=0.4\textwidth]{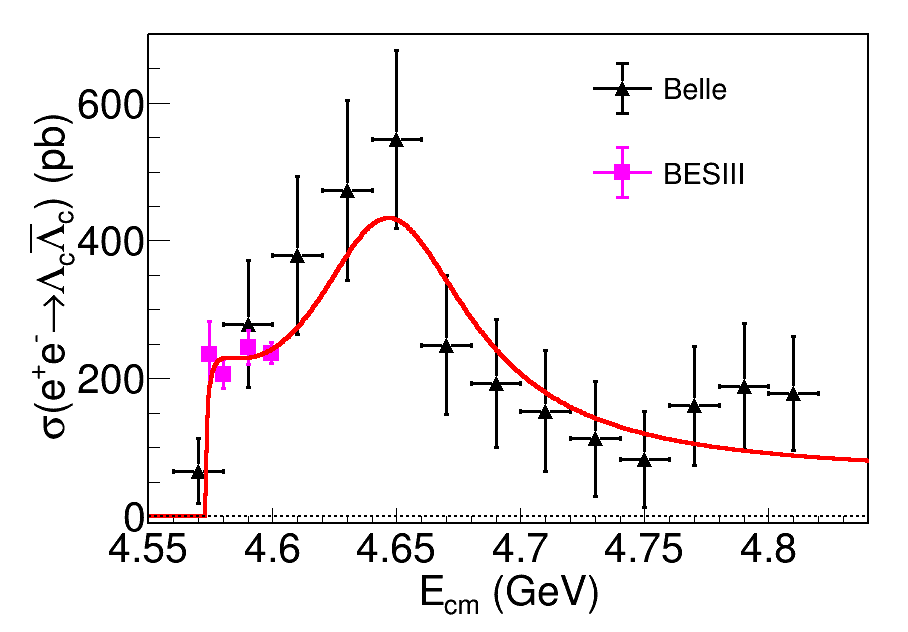}
\includegraphics[width=0.4\textwidth]{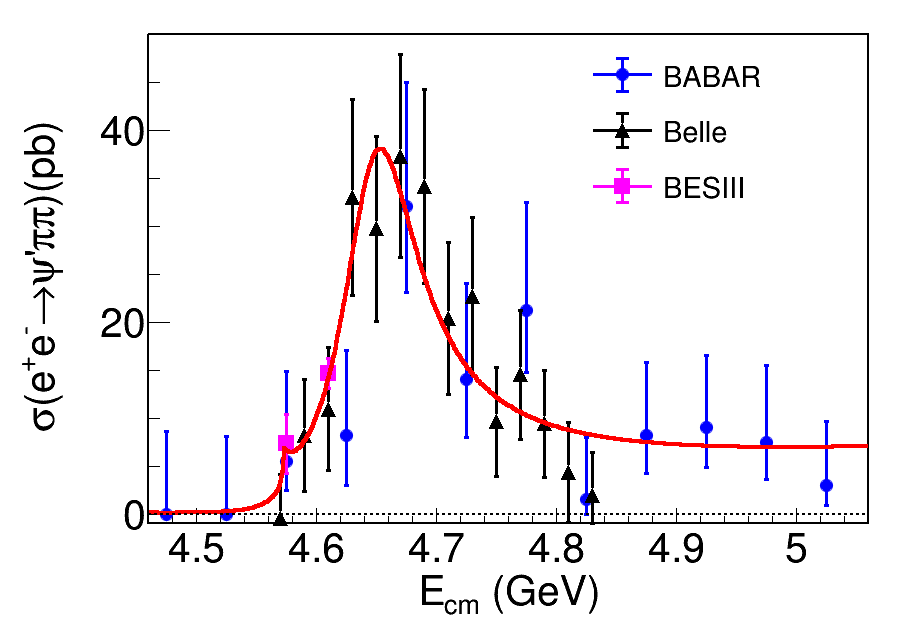}
\caption{\label{fig:fitc} The fit (red solid curves) to data of relevant processes with constant coherent background. }
\end{figure*}

Based on Eqs.~\eqref{intXLam}, \eqref{int4Lam} and \eqref{Vfdef}, a model of $K$ matrix type concerning a mixed mechanism of $s$ channel $X$ state and $\Lambda_c\bar \Lambda_c$ contact interaction, can be established as shown in Figs.~\ref{fig:KFSlam} and \ref{fig:KFSpsi}~\footnote{ The case with only FSI of $\Lambda_c\bar{\Lambda}_c$ is not considered in our model because it can not reproduce the $X(4660)$ peak in the fit. Moreover, the diagrams with $\gamma\to \Lambda_c\bar{\Lambda}_c$ vertices is considered as backgrounds since in those diagrams there are no bare $X(4660)$ propagators, causing a zero at $s=M_X^2$. }.
Specifically, the $K$ matrix sector satisfying final state theorem is
\begin{equation}\label{Km}
\begin{split}
&A_K=\frac{1}{1-i\rho K}=\frac{1}{1-i\rho(T_X+T_{ci})},\ \\
&\rho=\sqrt{\frac{s-4m^2}{s}},\
\end{split}
\end{equation}
where $T_X$ and $T_{ci}$ label the tree diagrams of $s$ channel $\Lambda_c\bar{\Lambda}_c\to X(4660)\to \Lambda_c\bar{\Lambda}_c$, and $\Lambda_c\bar{\Lambda}_c$ contact vertex as FSI in $^3S_1$ channel, respectively, see Fig.~\ref{fig:KFSlam}; $m$ is the mass of $\Lambda_c$. Note that other light channels, such as $\psi'\pi\pi$, do not show up in the above $K$ matrix, instead $\psi'\pi\pi$ only appears as final state. This greatly simplifies the calculation since it reduces the couple channel problem to a single channel approximation. This simplification is justifiable, as discussed in Ref.~\cite{Gong:2016hlt}, because the light channel thresholds are distant from the energy region near $X(4660)$, resulting in the replacement $K\to K+\Delta$ ($\Delta$ is a complex constant) in Eq.~\eqref{Km}. In fact $\text{Re}(\Delta)$ serves as a renormalization effect of the coupling constants, while $\text{Im}(\Delta)$ is small since the couplings to light channels are weaker. From a numerical point of view, this cannot significantly improve the fit quality. More importantly, the smallness of $g_2$ in Eq.~\eqref{Vfdef} is fully consistent with experimental observations [see Eq.~\eqref{Bran}]. Finally, in each channel we adopt a complex number serving as coherent background in the scattering amplitude.

Under the above formulations a combined fit to the data from both Belle~\cite{Wang:2014hta,Pakhlova:2008vn} and BESIII~\cite{Ablikim:2017lct,Ablikim:2017oaf} (also BABAR data~\cite{Lees:2012pv} for $\psi'\pi\pi$ final state) is performed, with a quite good fit quality $\chi^2/\text{d.o.f.}=26.6/33$, indicating that the present model is compatible with both Belle and BESIII data. As shown in Fig.~\ref{fig:fitc}, it is evident that the good fit quality originates from an enhancement near the $\Lambda_c\bar{\Lambda}_c$ threshold. Further investigations find that a virtual state lying below but very close to $\Lambda_c\bar{\Lambda}_c$ threshold, located at $M_V=4.566\pm0.007$ GeV, causes the enhancement\footnote{The fit with the light channel effects ($\Delta$) results in a local minimum with Im$(\Delta)=1.746\pm2.196$ (and the renormalized $\lambda$ becomes ($-17.606\pm3.258$)), but the quality $\chi^2/\text{d.o.f.}$ is not improved. The ``virtual state'' has gained an imaginary part: $M_V=4.566\pm0.003$ GeV and $\Gamma_V=0.002\pm0.002$ GeV. }. The main fit results are summarized in Table.~\ref{tab:fitres}.
\begin{table}[htbp]
\begin{center}
 \begin{tabular}  { c | c }
  \hline
  Parameters    &  Values\\
  \hline
  $M_{pole}$ (GeV) & $4.645\pm 0.028$\\
  $\Gamma_{pole}$ (GeV) & $0.078\pm 0.029$\\
  $M_V$ (GeV) & $4.566\pm0.007$\\
  $M_X$ (GeV) & $4.604\pm0.020$\\
  $g_1$ & $2.151\pm0.279$\\
  $\lambda$ (GeV$^{-2}$) & $-15.899\pm 2.718$\\
  \hline
 \end{tabular}\\
 \caption{Pole positions and fit parameters. $M_{pole}$ and $\Gamma_{pole}$ stand for the pole mass and width of $X(4660)$, respectively; $M_V$ is the position of the virtual state. The other parameters are those related to the poles: $M_X$ is the bare mass of $X(4660)$, $g_1$ is the coupling constant in Eq.~\eqref{intXLam}, and $\lambda\equiv 3C_V+C_A$, see Eq.~\eqref{int4Lam}.\label{tab:fitres}}
\end{center}
\end{table}

\begin{figure*}[t]
\centering
\includegraphics[width=0.4\textwidth]{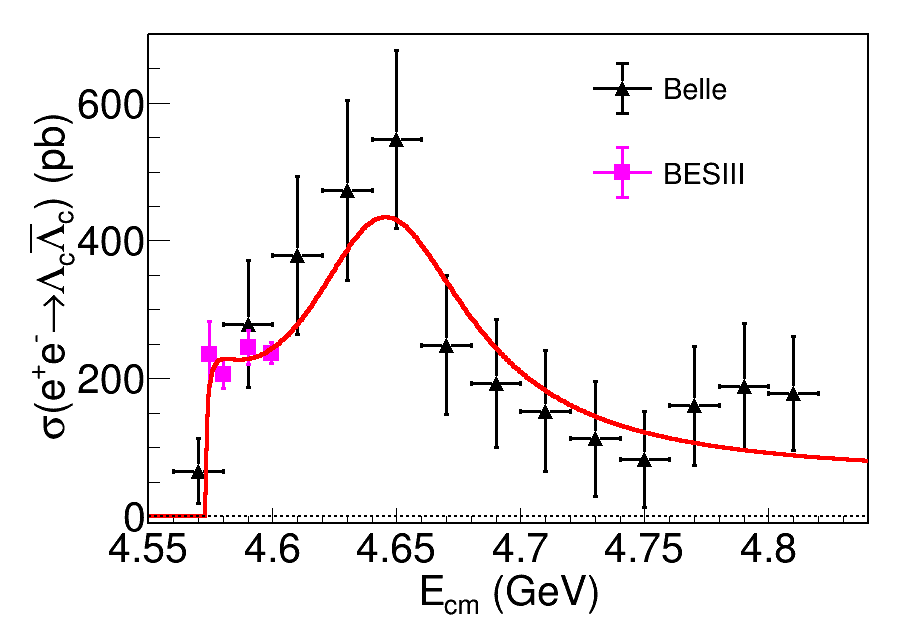}
\includegraphics[width=0.4\textwidth]{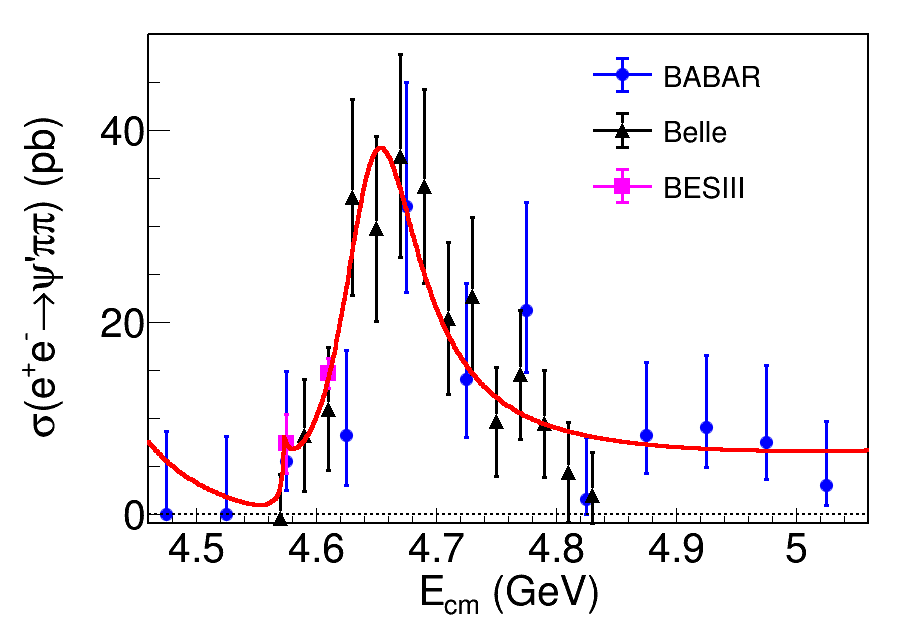}
\caption{\label{fig:fitbw} The fit (red solid curves) to data of relevant processes with explicit $X(4360)$ Breit-Wigner term. }
\end{figure*}

\begin{figure*}[htbp]
\centering
\includegraphics[width=0.45\textwidth]{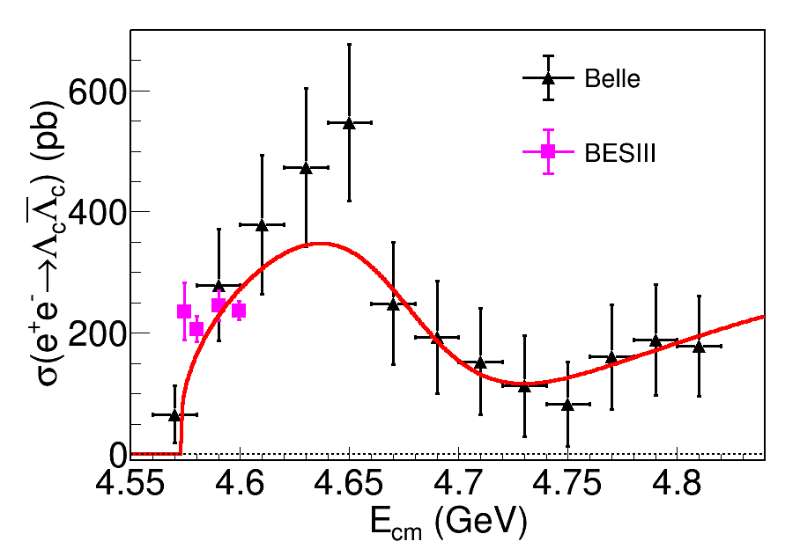}
\caption{\label{fig:Cofit} The fit (red solid curves) to data with constant coherent background plus Coulomb factor; the contact term generating the virtual state is absent. }
\end{figure*}

To proceed, the statistical significance of such virtual pole is studied. As a control, only Breit-Wigner effect is employed to fit all the data, giving the mass and width of $X(4660)$ as $M_{pole}=4.674\pm 0.043$ GeV and $\Gamma_{pole}=0.147\pm 0.110$ GeV, but the fit quality becomes $\chi^2/\text{d.o.f.}=44.4/34$. Comparing with the mixture mechanism ($\chi^2/\text{d.o.f.}=26.6/33$), the statistical significance of the virtual state is obtained to be $4.2\sigma$, indicating strong evidence in support of the virtual pole as truly existing. Furthermore, to test the stability of the poles, we also use the Breit - Wigner term for $X(4360)$ to replace the constant coherent background, with the mass and width fixed, while the residue varies. Even though the behaviour near $\psi'\pi\pi$ threshold changes a little (see Fig.~\ref{fig:fitbw}), the pole positions are found to be stable against the variation of backgrounds: the virtual state locates at $M_V=4.566\pm0.003$ GeV, and $X(4660)$ pole $M_{pole}=4.643\pm 0.011$ GeV and $\Gamma_{pole}=0.080\pm 0.019$ GeV. As is discussed at the beginning, Coulomb interaction plus $X(4660)$ Breit-Wigner cannot fit the data (especially the precision measurement results from BESIII) well within our expectation, see Fig.~\ref{fig:Cofit}. As a consequence, the Coulomb enhancement factor has little impact and can be neglected, which was also pointed out in Refs.~\cite{Baldini:2007qg,Haidenbauer:2016won}
.

\begin{figure*}[t]
\centering
\includegraphics[width=0.6\textwidth]{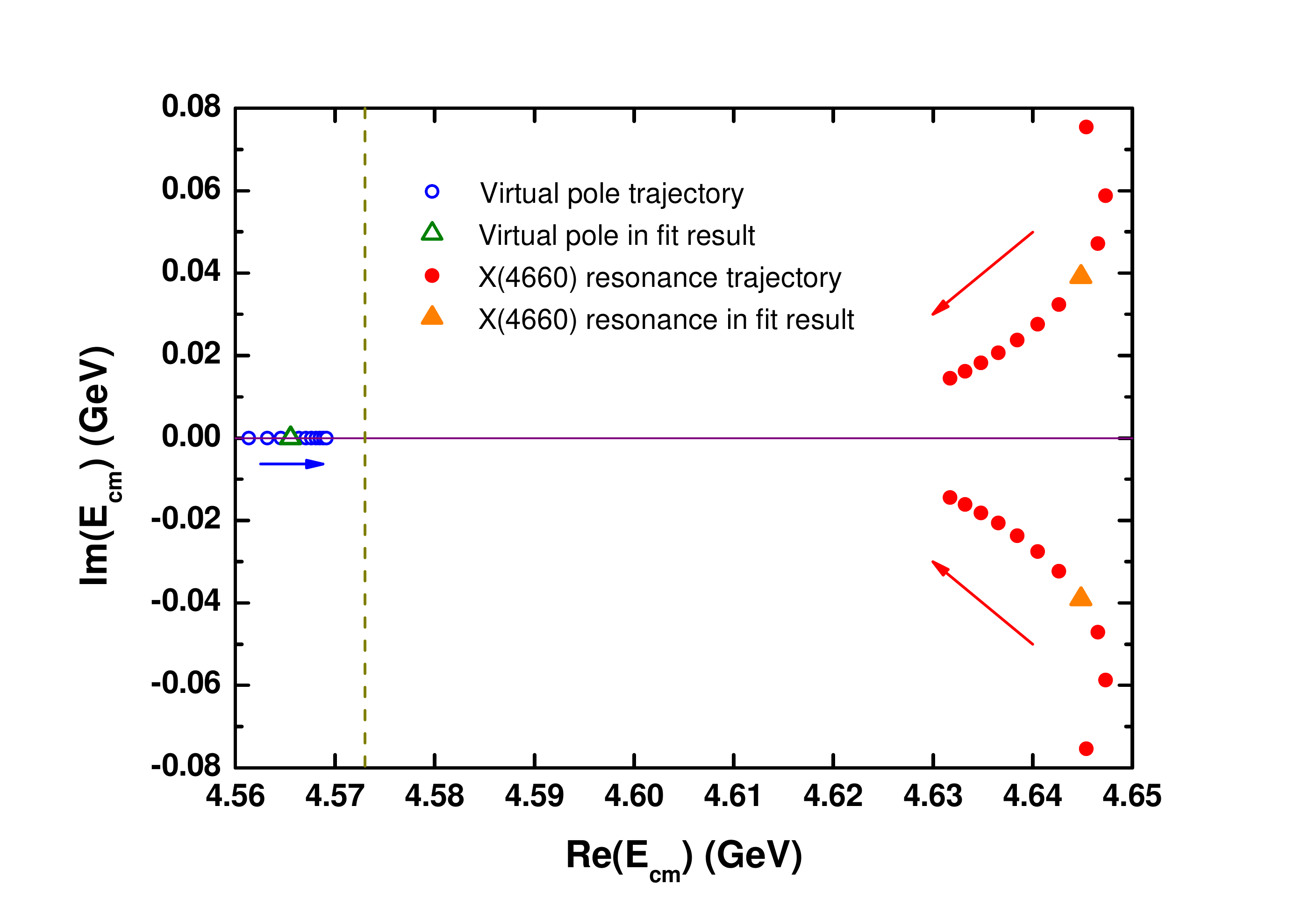}
\caption{\label{fig:pole} The trajectory of the poles in $E_{cm}\equiv\sqrt{s}$ plane against the increase of $|\lambda|$. The $|\lambda|$ value increases from $10$ GeV$^{-2}$ to $30$ GeV$^{-2}$ with the step $\Delta\lambda=2$ GeV$^{-2}$. The vertical dashed line labels the location of $\Lambda_c\bar{\Lambda}_c$ threshold. }
\end{figure*}

It should be emphasized that from a general point of view in quantum scattering theories, virtual states are believed to arise in attractive interactions that are not strong enough, being ``precursors'' of bound states: when the attractive coupling is strong enough they become bound states. In Table.~\ref{tab:fitres} the parameter $\lambda$ indicates the attractive force between $\Lambda_c$ and $\bar{\Lambda}_c$. Figure.~\ref{fig:pole} shows the trajectory of the poles against the increase of $|\lambda|$: when the $\Lambda_c\bar{\Lambda}_c$ contact interaction grows stronger, the virtual pole moves closer to the threshold; meanwhile, the $X(4660)$ resonance becomes narrower. It is worth noticing that the pole trajectory is model dependent: if the kinematic factor $i\rho$ were replaced by the entire two point function $B_0$, the virtual pole would go up to the first sheet and become a bound state, given a large enough $|\lambda|$. All in all, these analyses exhibits a very clear physical picture: the virtual pole is produced by FSI, while $X(4660)$ state is a typical Breit-Wigner state with a pair of nearby poles. According to the pole counting rule~\cite{Morgan:1992ge} (which has been successfully applied to the studies of ``$XYZ$'' physics in Refs.~\cite{Zhang:2009bv,Dai:2012pb,Gong:2016hlt,Gong:2016jzb}), our analysis suggests that $X(4660)$ is of confinement nature. Finally, notice that the virtual state should have an imaginary part when the light channels are taken into account, but its impact on the line-shape does not differ much from the present scheme.

Furthermore, the ratio between the decay widths $\Gamma(X\to \Lambda_c\bar{\Lambda}_c)$ and $\Gamma(X\to \psi'\pi\pi)$ can also be estimated as:
\begin{equation}\label{Bran}
\frac{\Gamma(X\to \Lambda_c\bar{\Lambda}_c)}{\Gamma(X\to \psi'\pi\pi)}\simeq \frac{\sigma(e^+e^-\to X\to \Lambda_c\bar{\Lambda}_c)}{\sigma(e^+e^-\to X\to \psi'\pi\pi)}\simeq 23,
\end{equation}
which is in agreement with Ref.~\cite{Cotugno:2009ys}.

In summary, this paper demonstrates the evidence of a virtual state in $e^+e^-\to X(4660)\to \Lambda_c\bar{\Lambda}_c$ process with significance of $4.2\sigma$. By employing a model with both $s$ channel $X(4660)$ propagator and $\Lambda_c\bar{\Lambda}_c$ FSI, the data from Belle and BESIII of $e^+e^-\to X(4660)\to (\Lambda_c\bar{\Lambda}_c)$ can be fitted well simultaneously. The virtual state plays a crucial role in respect to that fit since it gives a significant threshold enhancement. This pole is regarded as a $\Lambda_c\bar{\Lambda}_c$ molecular virtual state and would become a bound state if the $\Lambda_c\bar{\Lambda}_c$ contact coupling were larger, while $X(4660)$ is of confinement nature. Finally, since the statistics of the data is limited, the confirmation of it urgently recalls more experimental measurements with higher precisions, such as Belle II, BESIII, LHCb, etc.

{\it Acknowledgments.} We would like to thank Guang-Yi Tang, Xian-Wei Kang and R. Baldini Ferroli for helpful discussions and advice. This work is supported in part by National Nature Science Foundations of China (NSFC) under Contracts No. 10925522, No. 11021092.

\end{document}